\documentclass[11pt]{article} 
\usepackage{amsmath}
\usepackage{amsfonts,amssymb}

\begin{document}

\newcommand{\nl}{\nonumber\\}
\newcommand{\nnl}{\nl[6mm]}
\newcommand{\nle}{\nl[-2.5mm]\\[-2.5mm]}
\newcommand{\nlb}[1]{\nl[-2.0mm]\label{#1}\\[-2.0mm]}
\newcommand{\ab}{\allowbreak}

\newcommand{\e}{{\mathbf e}}
\newcommand{\mm}{{\mathbf m}}
\newcommand{\nn}{{\mathbf n}}

\newcommand{\be}{\bes}
\newcommand{\ee}{\ees}
\newcommand{\bes}{\begin{eqnarray}}
\newcommand{\ees}{\end{eqnarray}}
\newcommand{\eens}{\nonumber\end{eqnarray}}

\renewcommand{\d}{\partial}
\newcommand{\eps}{\epsilon}

\newcommand{\vect}{{\mathfrak{vect}}}
\newcommand{\map}{{\mathfrak{map}}}

\newcommand{\im}{{\rm im}\ }
\newcommand{\ext}{{\rm ext}\ }
\renewcommand{\div}{{\rm div}}

\newcommand{\dmu}{{\d_\mu}}
\newcommand{\dnu}{{\d_\nu}}

\newcommand{\tr}{{\rm tr}}
\newcommand{\oj}{{\mathfrak g}}
\newcommand{\hh}{{\mathfrak h}}

\newcommand{\J}{{\cal J}}
                                            
\newcommand{\TT}{{\mathbb T}}
\newcommand{\RR}{{\mathbb R}}
\newcommand{\CC}{{\mathbb C}}
\newcommand{\ZZ}{{\mathbb Z}}
\newcommand{\NN}{{\mathbb N}}

\title{{Why the Mickelsson-Faddeev algebra lacks unitary representations}}

\author{T. A. Larsson \\
Vanadisv\"agen 29, S-113 23 Stockholm, Sweden\\
email: thomas.larsson@hdd.se}

\maketitle 
\begin{abstract} 
A simple plausibility argument is given.
\end{abstract}

\vskip 1cm

Let $\oj$ be a finite-dimensional Lie algebra with generators $J^a$
and structure constants $f_{ab}{}^c$. The brackets are given by
$[J^a,J^b] = f^{ab}{}_c J^c$.
Denote the symmetric Killing metric (proportional to the quadratic 
Casimir operator) by $\delta^{ab} = \tr\ J^aJ^b$, and 
let the totally symmetric third Casimir operator be
$d^{abc} = \tr\ \{J^a, J^b\}J^c$. 

The current algebra $\map(M_3,\oj)$ is the algebra of maps from a
$3$-dimensional manifold $M_3$ to $\oj$. In local coordinates, the
generators are $\J_X = \int d^3x\ X_a(x) J^a$, where $X_a(x)$ are
functions on $\RR^3$. Define $[X,Y] = f^{ab}{}_cX_aY_bJ^c$.
This algebra admits an abelian extension known as the Mickelsson-Faddeev
(MF) algebra \cite{Fa84,Mi85,Mi89},
\bes
[\J_X,\J_Y] &=& \J_{[X,Y]} + d^{abc}\int d^3x\ 
\eps^{\mu\nu\rho} \dmu X_a(x) \dnu Y_b(x) A_{c\rho}(x), \nl
{[}\J_X, A_{a\mu}(x)] &=&  f^{bc}{}_a X_b(x) A_{c\mu}(x) + \dmu X_a(x), 
\label{MF1}\\
{[}A_{a\mu}(x), A_{b\nu}(y)] &=& 0,
\eens
where $\eps^{\mu\nu\rho}$ is the totally
anti-symmetric epsilon tensor in three dimensions.
If we specialize to the $3$-torus $\TT^3$, we can expand all fields in 
a Fourier basis. The algebra $\map(\TT^3,\oj)$ then takes the form
\bes
[J^a(\mm), J^b(\nn)] &=& f^{ab}{}_c J^c(\mm+\nn)
  + d^{abc} \epsilon^{\mu\nu\rho} m_\mu n_\nu A_{c\rho}(\mm+\nn), \nl
{[}J^a(\mm), A_{b\nu}(\nn)] &=& -f^{ac}{}_b A_{c\nu}(\mm+\nn) 
  + \delta^a_b m_\nu \delta(\mm+\nn), 
\label{MF2}\\
{[}A_{a\mu}(\mm), A_{b\nu}(\nn)] &=& 0.
\eens
Here $\mm = (m_\mu) \in \ZZ^3$ is a momentum labelling the Fourier
modes, and $J^a(\mm)$ and $A_{a\mu}(\mm)$ are the Fourier components of
the algebra generators and the gauge connection, respectively.

No physically relevant representations of the MF algebra are known, and
indeed a kind of no-go theorem was given by Pickrell long ago
\cite{Pic89}: the algebra (\ref{MF1}) has no faithful, unitary
representations on a separable Hilbert space. The purpose of the present
note is to give a very simple argument why this must be true. The idea
is to consider the restrictions of the torus algebra (\ref{MF2}) to
various loop algebras. The restriction of a unitary representation to
any subalgebra is obviously still unitary. However, it is well known
that the only unitary representation of a proper loop algebra is the
trivial representation. Since all restrictions of a unitary torus
algebra representation to its loop subalgebras are trivial, the torus
algebra representation must itself be trivial.

Let $\e = (e_\mu)$ be a vector in $\ZZ^3$. A loop subalgebra is
generated by elements of the form
\be
J^a_m = J^a(m\e).
\label{restrict}
\ee
It is straightforward to verify that
\be
[J^a_m, J^b_n] = f^{ab}{}_c J^c_{m+n},
\label{loop}
\ee
i.e. the restriction of the MF extension to this subalgebra vanishes.
The proof only uses anti-symmetry of the epsilon symbol, 
$\eps^{\mu\nu\rho}e_\mu e_\nu \equiv 0$.
The algebra (\ref{loop}) is recognized as a proper loop algebra, i.e. an
affine Kac-Moody algebra with zero central extension. It is well known
that all non-trivial lowest-energy unitary representations of affine 
algebras have a
positive central charge \cite{GO86}. Hence the restriction of a unitary
MF representation to this subalgebra is trivial. Since this must be true
for every choice of vector $\e$, we conclude that the unitary 
representation of the MF algebra must itself be trivial.

The argument has one loophole: the loop algebra representation is
supposed to be of lowest-energy type. The algebra (\ref{loop}) certainly
has unitary representations if we relax this condition, e.g. the direct
sum of one lowest- and one higher-energy unitary representation with
opposite central charges, or classical representations on fields
valued in $\oj$ modules. However, those are not the kind of
representations that we expect in quantum systems, where there should be
a Hamiltonian which is bounded from below. This Hamiltonian induces a
grading by energy of every loop subalgebra. Hence the restriction to
every such subalgebra should be of lowest-energy type, and some should
be non-trivial.

Mickelsson has studied another type of representation, where the algebra
has a natural and fiberwise unitary action in the bundle of fermionic
Fock spaces parametrized by external vector potentials \cite{Mi92}. In
other words, these representations describe quantum chiral fermions
living over a classical background gauge field. Whereas this
construction is mathematically interesting, it can not fundamentally
describe physics, where the gauge fields must be quantized as well.

This result implies that conventional gauge anomalies proportional to
the third Casimir operator are inconsistent. Namely, the gauge generators
must be represented by unitary operators. However, we have just seen 
that this means that the representation is trivial. Since the MF
extension vanishes in the trivial representation, the anomaly must 
indeed be zero. This result is of course consistent with physical
intuition \cite{Bon86,NAG85}. 

The current algebra $\map(\TT^N,\oj)$ also admits another extension,
first found by Kassel \cite{Kas85}. It is usually called the central
extension, although the extension does not commute with diffeomorphisms.
In a Fourier basis, this extension is defined by the brackets
\bes
[J^a(\mm), J^b(\nn)] &=& f^{ab}{}_c J^c(\mm+\nn)
  + k \delta^{ab} m_\rho S^\rho(\mm+\nn), \nl
{[}J^a(\mm), S^\mu(\nn)] &=& [S^\mu(\mm), S^\nu(\nn)] = 0, 
\label{central}\\
m_\mu S^\mu(\mm) &\equiv& 0.
\eens
The restriction to the subalgebra generated by (\ref{restrict}) reads
\be
[J^a_m, J^b_n] &=& f^{ab}{}_c J^c_{m+n} + k \delta^{ab} m \delta_{m+n}, 
\label{KM}
\ee
where $S_m \equiv e_\mu S^\mu(m\e) \propto \delta_m$,
because the condition $m S_m = 0$ implies that $S_m$ is proportional to
the Kronecker delta $\delta_m$. Equation (\ref{KM}) is recognized 
as an affine Kac-Moody algebra, including the central term. Since the
Kac-Moody algebra has unitary representations for positive central
charge, the argument above does not apply to the algebra
(\ref{central}). Nothing prevents it from having unitary, lowest-energy
representations, and hence such gauge anomalies may occur in physics. In
fact, lowest-energy representations of the algebra (\ref{central}) 
have been known 
since 1990 \cite{MEY90}, and it was recently shown that this kind of 
gauge anomaly does arise when one quantizes
the observer's trajectory together with the fields \cite{Lar05a}.

To conclude, we observed that a lowest-energy representation of a torus
algebra can only be unitary if all restrictions to loop algebras are so,
and that unitarity of loop algebra representations requires an extension
proportional to the quadratic Casimir. This rules out the MF extension,
because it is proportional to the third Casimir. The Kassel extension
can have unitary representations. The result were formulated on
the three-dimensional torus for convenience, but this not a critical
assumption. On a general manifold, we can consider the restrictions to
elementary loops instead; the number of such loops is given by the first
Betti number.

\medskip
I thank Jouko Mickelsson for pointing out a weakness in the original
argument.

\end{document}